\documentclass[conference]{IEEEtran}

\makeatletter
\def\ps@headings{%
\def\@oddhead{\mbox{}\scriptsize\rightmark \hfil \thepage}%
\def\@evenhead{\scriptsize\thepage \hfil \leftmark\mbox{}}%
\def\@oddfoot{}%
\def\@evenfoot{}}
\makeatother

\usepackage{cite}
\usepackage{url}
\usepackage[dvips]{graphicx}
\usepackage{color,subfigure,url,amsmath}

\newcommand{\nael}[1]{\color{red}{ \em #1 }\color{black}}

 \renewcommand{\nael}[1]{}

\begin{document}
\title{How do Wireless Chains Behave?\\ The Impact of MAC Interactions}
\author{\IEEEauthorblockN{Saquib Razak}
\IEEEauthorblockA{School of Computer Science\\
Carnegie Mellon University\\
srazak@cmu.edu}
\and
\IEEEauthorblockN{Vinay Kolar}
\IEEEauthorblockA{Dept. of Wireless Networks,\\
RWTH-Aachen University\\
vko@mobnets.rwth-aachen.de}
\and
\IEEEauthorblockN{Nael B. Abu-Ghazaleh}
\IEEEauthorblockA{
Carnegie Mellon University\\
and
SUNY, Binghamton\\
nael@cs.binghamton.edu}
\and

\IEEEauthorblockN{Khaled A. Harras}
\IEEEauthorblockA{School of Computer Science\\
Carnegie Mellon University\\
kharras@cs.cmu.edu}}

\maketitle

\begin{abstract}

In a Multi-hop Wireless Networks (MHWN), packets are routed between
source and destination using a {\em chain} of intermediate nodes;
chains are a fundamental communication structure in MHWNs whose
behavior must be understood to enable building effective protocols.
The behavior of chains is determined by a number of complex and
interdependent processes that arise as the sources of different chain
hops compete to transmit their packets on the shared medium.  In this paper, we show
that MAC level interactions play the primary role in determining the
behavior of chains.  We evaluate the types of chains that occur based
on the MAC interactions between different links using realistic
propagation and packet forwarding models.  We discover that the
presence of destructive interactions, due to different forms of hidden
terminals, does not impact the throughput of an isolated chain
significantly.  However, due to the increased number of
retransmissions required, the amount of bandwidth consumed is
significantly higher in chains exhibiting destructive interactions,
substantially influencing the overall network performance.  These
results are validated by testbed experiments.  We finally study how
different types of chains interfere with each other and discover that
well behaved chains in terms of self-interference are more resilient
to interference from other chains.


\end{abstract}

\section{Introduction}
\label{sec:intro}
Multi-Hop Wireless Networks (MHWNs), which include mesh, sensor, and
ad hoc networks, are forecast to play an important role in an Internet
that will grow increasingly wireless at the edge.  MHWNs reduce
infrastructure requirements by having wireless nodes relay traffic
towards access points; they are attractive whenever infrastructure is
unavailable or costly, or quick deployment is
desired~\cite{ref:india_net,ref:MANET_2,ref:roofnet_arch_eval}.  The
complex and dynamic nature of wireless propagation, interference, and
user mobility make developing effective networking protocols for MHWNs
a significant challenge.

In an MHWN, packets are forwarded from source to destination using a
{\em chain} of nodes.  Starting from the source, a node forwards
packets to the next node in the chain forming a {\em path} towards the
destination.  Chains represent a fundamental communication structure
in MHWNs, and understanding their behavior is critical to designing
effective protocols.  In particular, routing protocols must discover
efficient chains that can then be used for communication.  Early
routing protocols used path length to discriminate between chains,
favoring the shortest available
path~\cite{perkins-94,ref:dsr,ref:perkins_03_AODV,ref:olsr}.
Recently, individual link qualities have been taken into account in
evaluating path quality~\cite{ref:lqsr}.  However, the behavior of a
chain is a complex process that cannot be accurately characterized by
looking at the individual links without consideration to how they
interact with each other.

Several studies have examined the behavior of chains.  Li et al. study
the performance of chains as the number of hops are
increased~\cite{ref:li_chain}.  They also study the effect of
cross-interference between chains.  Xu and Sadaawi analyze TCP
instability due to chain self-interference and discover short-term and
long-term unfairness issues in cross-chain
interactions~\cite{ref:xu_02_macTcp}. Ping et al. present the effect of
traffic on routing instability, packet drops, and unfairness due to
self interference within chains~\cite{ref:ping_thruput}.  These
analyses significantly differ from ours in a number of ways, including
the fact that they do not consider the detailed processes, such as the
impact of the MAC level interactions, on the performance of chains.
We review these and other related works in Section~\ref{sec:related}.

Several complex and inter-dependent processes combine to determine the
behavior of chains.  In particular, the performance of chains is
affected by {\em self-interference} among the different hops of 
chain as they compete to transmit on shared wireless medium.  This
interference not only reduces the available transmission time at each
hop, but also causes packet collisions due to a variety of {\em MAC
  level interactions} that occur in different chains.  Moreover,
nodes in the middle of the chain experience higher interference than
nodes at the edge because they are in interference range with more nodes in the chain; this is a process we call {\em contention
  unfairness}.

Among the different processes that impact chain performance, MAC level
interactions play a central role.  They also significantly moderate
the effect of the other processes. In order to better understand chain
behavior, we first analyze the types of interactions that occur most
frequently in four-hop chains.We later explore generalizing these
results.  Although there is a large number of potential interaction
configurations that may arise in chains, we discover that only a small
number of them occur in practice due to chain geometry restrictions.
Specifically, we set up forwarding rules to produce chains in a way
representative of how routing protocols work.  We use a Signal to
Interference and Noise Ratio (SINR) model for packet reception which
allows us to account for the effect of capture.  

We then evaluate, in Section~\ref{sec:performance}, the effects of the
interference interactions within a chain on its overall
performance. We first use simulation to determine the throughput and
the number of packet drops for different types of chains.  Afterwards,
we validate our simulation results by comparing them against results
obtained from an experimental testbed. 

 The next contribution of the paper, discussed in
 Section~\ref{sec:nhops}, is to develop an approach for estimating the
 performance of general n-hop chains.  Specifically, we observe that
 the best chains are those where senders are in carrier sense range
 with each other, allowing the MAC protocol to effectively arbitrate
 the medium.  In those chains where this is not the case, the presence
 of a hidden terminal has a higher impact than a hidden terminal with
 capture.  Finally, the location of the hidden terminal is also a
 factor in determining the performance; the earlier the hidden
 terminal, the worse its impact.


After characterizing how a single chain self-interferes in isolation,
we look at the problem of how multiple chains interfere with each
other.  In a general MHWNs, multiple connections are active
simultaneously, interfering with each other. In
Section~\ref{sec:crosschain}, we evaluate cross-chain interactions and
study their effects on the performance of chains.  Again, we discover
that the presence of hidden terminals significantly affects
performance and fairness.  Moreover, we discover that well behaved
chains in terms of self-interference are more resilient to destructive
interference from other chains.  Finally, we summarize our
contributions and present some concluding remarks in
Section~\ref{sec:conclusion}.

\section{Related Work}
\label{sec:related}



Several researchers have developed framework to characterize the behavior of wireless networks~\cite{ref:garetto_06_starve,ref:bianchi_00_802.11,ref:boorstyn_87_throughput}. They develop models to predict the behavior of networks but do not consider factors affecting chains in multi-hop networks. Works like MACA, MACAW and FAMA added MAC level packet exchange combined with Carrier Sensing to mitigate hidden terminal problems~\cite{ref:bharghavan_94_macaw,ref:karn_90_maca,ref:fullmer_97_ht}. These protocols do not solve the hidden terminal problems under realistic models for wireless interference and packet reception. 

Jain et al. model interference in a network as a conflict graph and estimate throughput achievable in a given network and load~\cite{ref:jain_03_interference}. They utilize global knowledge of network and interference to determine routes that maximize throughput. They do not analyze self interference between chain links and do not consider chain effects while evaluating these routes. 

Recently, Gollakot and Katabi present an interference cancellation
technique that uses information from successive collisions to decode
collided packets~\cite{ref:katabi_hiddenterminal}. Their techniques
assume symmetric hidden terminals that causes same packets to collide
several times. Garetto et al. and Razak et al. show that most
interactions in ad-hoc networks in general and chains in particular have asymmetric interference hence
this interference cancellation technique is not applicable in most
cases~\cite{ref:garetto_05_twoflow,ref:srazak_twoflow,ref:srazak_08_selfInterference}.

Li et al. studied the performance of chains as the number of hops are
increased~\cite{ref:li_chain}. They analyze the effect of MAC 802.11
behavior on the performance of multi-hop chains but do not categorize
interference patterns that govern network performance in terms of
throughput and bandwidth utilization. They also studied the effect of
cross-interference between chains.  Ping et al. present a hop by hop
analysis of a multi-hop chain and study the effects of hidden nodes on
the throughput of a chain topology~\cite{ref:ping_thruput}.  They
present a quantitative approach towards estimating the throughput of a
chain. They provide two main observations about flows in a
chain. Firstly the presence of hidden nodes cause packet drops that
reduce the throughput of the chain directly, and secondly packet drops
cause reporting of broken links to the routing protocol and hence
reducing the throughput indirectly.

In earlier work, Razak et al. use a simplified two-disc binary model of packet
reception to categorize interactions between self interfering links of
a chain~\cite{ref:srazak_08_selfInterference}. 
The current paper advances this earlier study in several important
ways: (1) Enumerates factors that are instrumental in affecting chain behavior; 
(2) it uses the SINR propagation model, which allows us to more
accurately model interactions, and include the important impact of
capture; (3) it presents experimental validation of the results; (4)
it contributes a more accurate approach for estimating chain
probabilities taking into account the effect of routing protocols and
the node density; (5) it presents a generalization to n-hop chains;
and (6) it presents a study of interactions across chains.


In summary, most of the work that analyzes chains concentrates on observing the behavior of chains and then identifying and evaluating the effects that cause these behaviors. Our approach, studies the factors that determine chain behavior from first principles, identifies the factors that have high impact and then evaluates the effect of these factors on chain performance.

\section{MAC Interactions in Chains}
\label{sec:interactions}

In this section, we first discuss the different factors that impact
chain behavior.  We identify that MAC interactions between chain links
have the highest impact on chain performance.  We then study the
frequency of occurrence of the different sets of interactions in 4 hop
chains, under representative routing protocols.  

\subsection{Factors that Determine Chain Behavior}
\label{sec:determinants}
It is well known that the throughput of a chain decreases as the number of hops increase~\cite{ref:li_chain}. In this section we outline the factors that affect the performance of chains of a given length.

\noindent \textbf{Contention Unfairness:} As nodes in a chain compete
for channel access, the ones in the middle of the chain may contend
with more nodes within the chain than those at the edges. These
middle nodes have a smaller chance to transmit, which results in
longer packet queues, ultimately leading to packet drops. We term this
effect {\em contention unfairness} which affects the overall
performance of the chain and is similar to the {\em flow in the
middle}
problem~\cite{ref:wang_05_model}.

\noindent \textbf{MAC Level Interference Interactions:} One of the factors that
affects the performance of chains is the {\em types of MAC interaction}
between hops of the chains that do not share a common node.  For example, if the source of one hop is a hidden terminal to the receiver of another, the performance of the chain is significantly influenced.

\noindent \textbf{Pipeline Effect:} In a chain, earlier hops feed data to later ones. As a result, later hops can never transmit more
packets than earlier ones.  This effect has important implications: if
there is an interaction leading to unfairness in favor of later hops,
it cannot be sustained.  On the other hand, unfairness in favor of
earlier hops leads to rate mismatch at intermediate hops and packet
drops in queues.  This effect moderates the impact of hidden terminals
and contention unfairness, preventing unfairness in some cases.

\noindent \textbf{Cross-chain Interference:} Chains do not exist in isolation within a network. Links in a chain can have different interactions with links within other chains affecting the overall performance of the network. The effect of this {\em cross chain interference} is an important factor to consider while characterizing the performance of chains.

Of the factors above, the MAC level interactions play a defining role
in the overall performance of the chain. Contention Unfairness,
Pipelining Effect and Cross-chain interference may exist in all chains
to varying degrees but the MAC interactions significantly influence
the impact that these other effects have.  We show examples of this
behavior in Section~\ref{sec:crosschain}. Thus, the first and most
important step in understanding chains is to understand the occurrence
probability and impact of the different MAC interactions within
chains.


\subsection{MAC interactions between two links}


In this section, we introduce MAC level interactions in the context of
two interfering links~\cite{ref:garetto_05_twoflow,ref:twoflow_sinr}.
This scenario is the simplest case in which links interfere at the MAC
level.  It provides a basis for identifying the different interaction
cases, which we then use to classify interactions that occur within
chains.

In a wireless network, the state of the channel at the receiver
determines whether a reception occurs successfully or not. However,
carrier sensing is carried out at the sender in Carrier Sense Multiple
Access (CSMA) protocols.  Accordingly, if the receiver channel is busy
but the sender channel appears idle, a collision can occur.  The
geometry of the interfering links (more accurately, the state of the
channels between them) determines what MAC level interactions arise.
These interactions can significantly impact performance or cause short
term or long term unfairness.

Given two interfering links $S1-D1$ and $S2-D2$, the type of MAC
interaction that occurs depends on the state of the secondary (or
unintended) channels between $S1-S2$, $S1-D2$ and $D1-D2$.  Each of
these channels can be in a number of states (in reception range, in
carrier sense range, in interference range, or in interference range
with capture), resulting in a large number of interaction
types~\cite{ref:garetto_05_twoflow,ref:twoflow_sinr}. Garetto
\textit{et al.} identify 5 categories of interactions in a simplified
unit-disc model of interference~\cite{ref:garetto_05_twoflow}. Razak
\textit{et al.} identify 10 categories of basic interactions under
SINR model. We next summarize the three interaction categories that
occur most frequently in chains~\cite{ref:twoflow_sinr}.


\begin{figure}
 \centering
\includegraphics[scale=0.15]{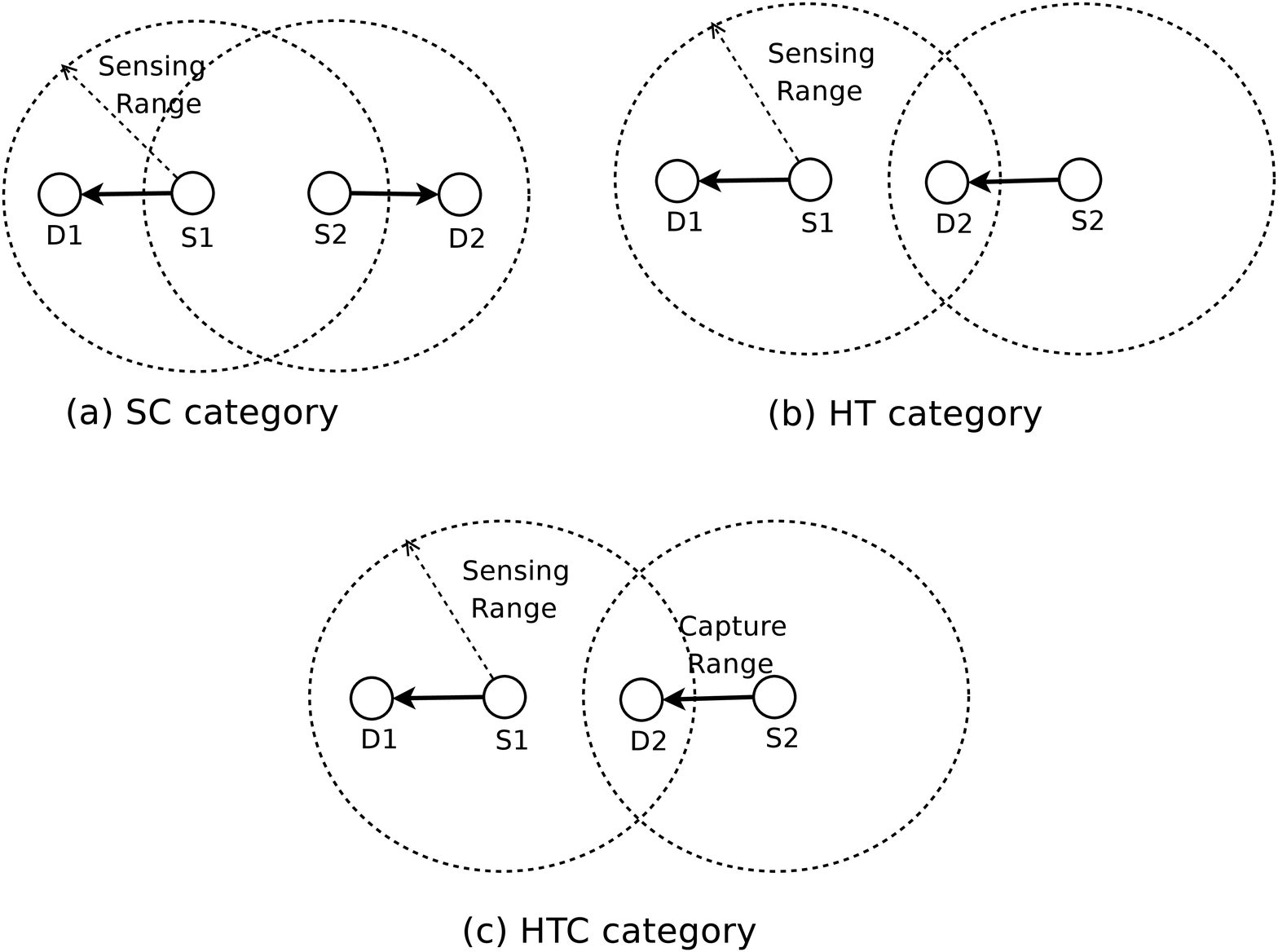}
\caption{Categories of link interaction.}
\label{fig:linkInteraction}
\end{figure}
\noindent \textbf{1. Senders Connected Symmetric Interference (SCSI or
  SC):} SCSI includes all scenarios where the sources of the two links
can sense each other (Figure~\ref{fig:linkInteraction}(a)). Thus, CSMA
prevents senders from concurrent transmissions; and no collisions
other than those arising when the two senders start transmission at
the same time will occur (not giving CSMA a chance to work). Such
collisions are unavoidable, and their probability is low due to the
randomization of the backoff period. We will henceforth refer to the
SCSI interaction as \textit{SC} for simplicity.

\noindent \textbf{2. Asymmetric Incomplete State (AIS or HT):} The
senders are not connected in AIS scenarios and, hence, can transmit
concurrently. Each sender has incomplete information about the state
at the respective receivers. As shown in
Figure~\ref{fig:linkInteraction}(b), an asymmetric interference is
observed where a transmission from the one sender $S_1$ causes packet
collision at the receiver $D_2$ of the other link. The link $S_1D_1$
is unaffected by signals from $S_2$ to $D_2$. The source $S2$ observes
large backoff values due to repeated packet collision and hence the
throughput of $S_2D_2$ is significantly affected. For simplicity, we
henceforth refer to the AIS interaction as \textit{HT} since it
experiences severe Hidden-Terminal effect.

\noindent \textbf{3. Hidden Terminal with Capture Effect (HTC):} In this
interaction, two links have {\em HT} interaction but the destination
with the hidden terminal problem is able to capture its packets from
its source under interference from the opposite
source. Figure~\ref{fig:linkInteraction}(c) shows one possible
placement of nodes with HTC interaction. In this case, although $D_2$
is in interference range of $S_1$, it is able to capture its packets
from $S_2$ as long as the packet from $S_2$ arrives at $D_2$ before
$S_1$ starts transmission. Recent studies have shown that a node can
capture packets if it has locked on to the packet before the
interfering nodes starts transmitting~\cite{ref:kochut_capture}. If
the interfering node starts transmitting first, the destination node
will lock on to its signal and will not be able to decode the packet.

While other categories exist (for example, symmetric hidden terminals
where both packets are lost), we show in the next section that they
almost never arise in chains due to the geometric structure of chains
selected by a forwarding rule representative of MHWN routing protocols.

\subsection{What interactions occur most frequently in chains?}

In this section, we determine the probabilities of different types of
interactions that occur between links in multi-hop chains.  We start
with a uniform deployment of the nodes in a fixed-size area.  We
considered using shortest path routing to select the chains.  However,
since modern routing protocols incorporate link quality in evaluating
paths, we decided to use the following forwarding rule instead to
generate paths.  We start from the source and pick as the next hop the
neighbor that is expected bring the packet the closest to the
destination taking into account both distance and link quality.  This
expected distance is the product of the actual distance travelled
towards the destination divided by the expected number of
retransmissions necessary to delivery the packet
(ETX~\cite{ref:decouto_03_metric}).  Link qualities were assumed to be
distributed as a function of distance between the sender and receiver
according to the log-normal shadowing distribution.  This forwarding
rule is identical to that implemented by the NADV routing
protocol~\cite{ref:nadv}.

\begin{figure}[ht]
\begin{center}
\includegraphics[scale=0.35]{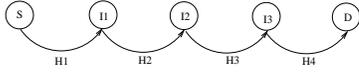}
\caption{A Chain with 4 hops.}
\label{fig:chain4hops}
\end{center}
\end{figure}

We conduct our analysis of interactions on chains with 4 hops. We
choose 4-hop chains because they have multiple interactions between
their links and provide insights that are helpful in generalizing our
evaluation to n-hop chains as we will show in
Section~\ref{sec:nhops}. In a four hop chain as depicted in
Figure~\ref{fig:chain4hops}, there are three different sets of links
that can be active at the same time. These sets of links result in
four-hop chains exhibiting three types of interference
interactions. We denote this set of interactions as INT1/INT2/INT3,
where INT1 represents interaction between hops H1 and H4, INT2
represents interaction between H1 and H3, and INT3 represents
interaction between hop H2 and hop H4.

\begin{figure*}[htp]
\begin{center}
        \mbox{
        \subfigure[Legend.\label{fig:legend}]{\includegraphics[scale=0.40]{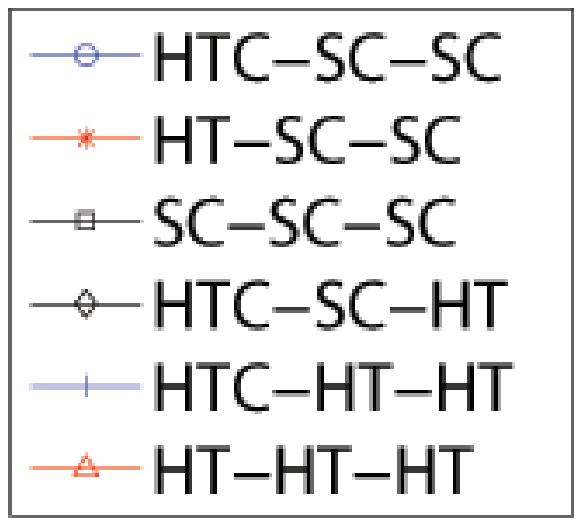}} 
        }
	\mbox{
        \subfigure[Number of Nodes = 225 nodes.\label{fig:den225}]{\includegraphics[scale=0.30]{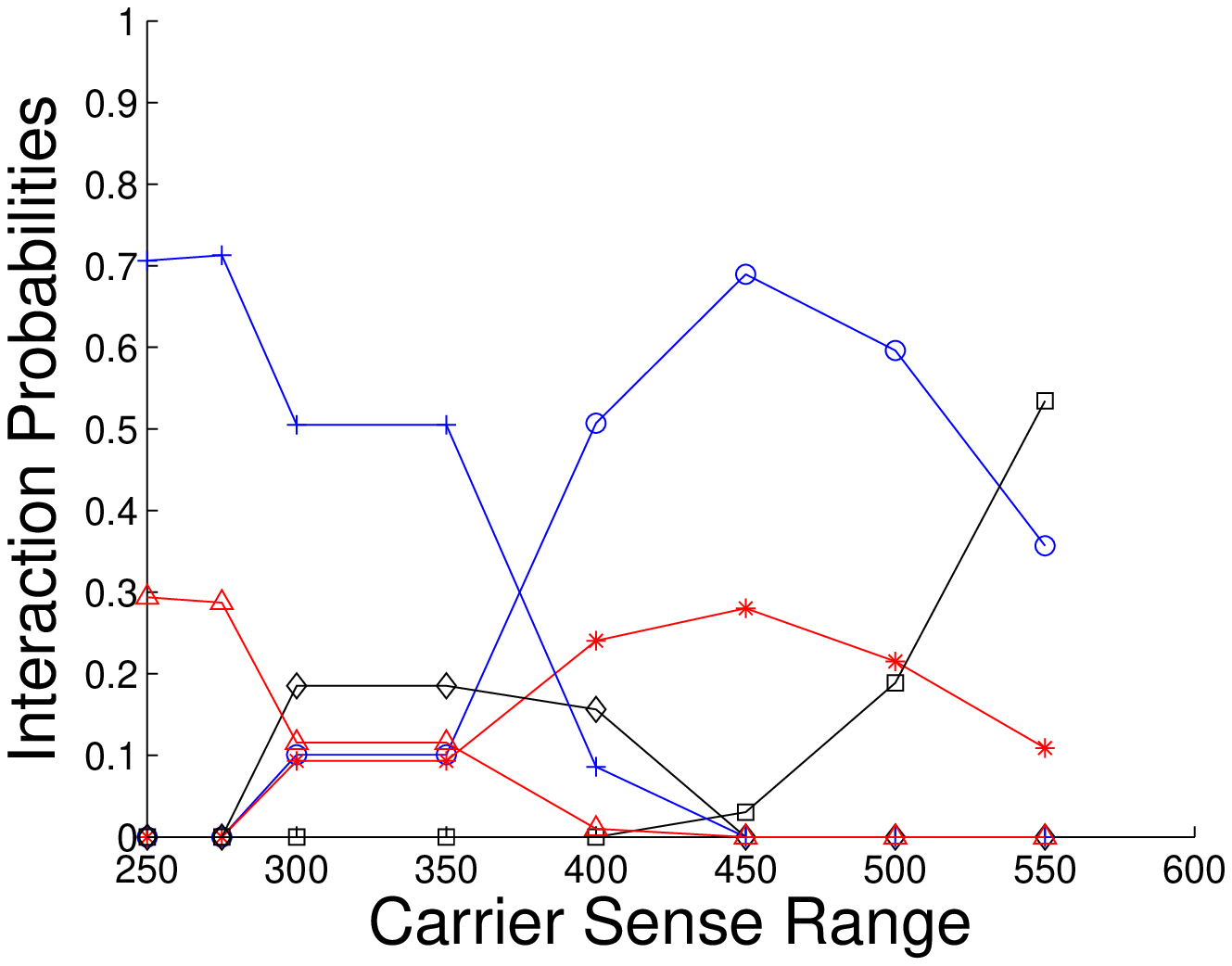}} 
        }
        \mbox{
        \subfigure[Number of Nodes = 500 nodes.\label{fig:den500}]{\includegraphics[scale=0.30]{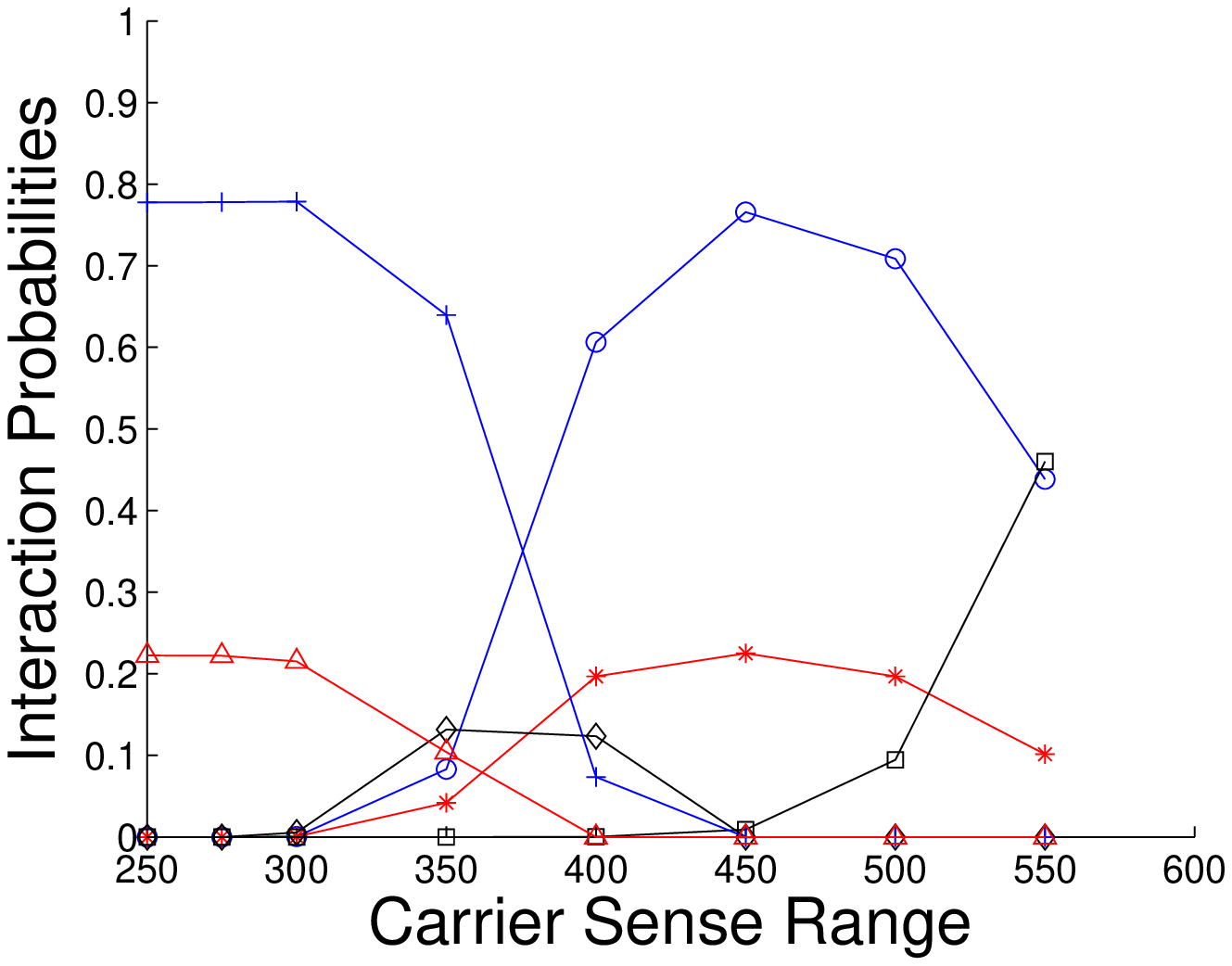}} 
        }
        \mbox{
        \subfigure[Number of Nodes = 900 nodes\label{fig:den900}]{\includegraphics[scale=0.30]{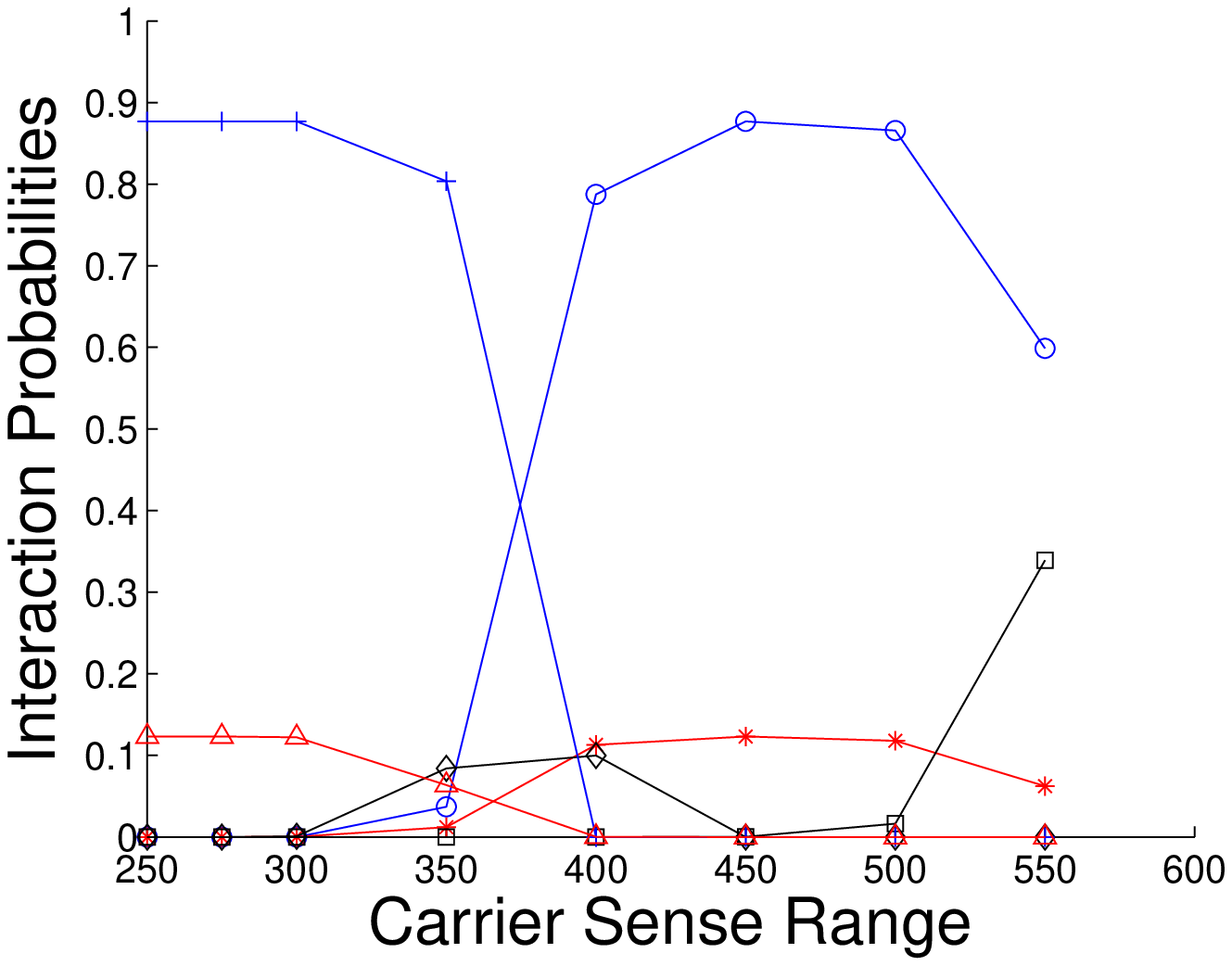}} 
        }
\end{center}
\caption{Probabilities of Interactions between links of a four-hop chain in a 1500m x 1500m network. }
\label{fig:probabilities}
\end{figure*}

In order to evaluate routes picked by NADV-like forwarding rule, we
generate a topology with nodes uniformly distributed in a 1500 x 1500
meters area. We study the impact of node density by increasing the
number of nodes deployed in the same area.  We observe that the
interaction probabilities stabilize as density of nodes increases and
are not significantly different at lower densities. Next we calculate
the route from each node to every other node using the forwarding rule
and evaluate the interference interactions between links of all 4-hop
routes. Figure~\ref{fig:probabilities} shows the probability of the
different types of interactions in 4-hop chains for different node
densities.  In this Figure, we omit some very rare interactions that
occurred to avoid clutter.

As shown in Figure~\ref{fig:probabilities}, in sparse networks,
routing protocols are forced to pick longer hops, leading to a higher
percentage of hops with Hidden Terminal interactions. Given a density,
the occurrence of interactions are a function of Carrier Sense range,
as we have more senders connected with higher Carrier Sense range. As
we decrease the Carrier Sense range, more nodes can transmit together
causing a higher number of hidden terminal problem. We observe from
this figure that there is a substantial number of hidden terminal
interactions for all values of Carrier Sense range. For values of
Carrier Sense range representative of those used on commercial
wireless cards, interactions SC/SC/SC, HT/SC/SC, and HTC/SC/SC occur
most often. 

\section{Chain Performance}
\label{sec:performance}
In this section we evaluate the performance of 4-hop chains with different interference interactions. We pick interactions: SC/SC/SC, HTC/SC/SC, and HT/SC/SC since they occur most commonly at  realistic values for Carrier Sense range. We evaluate the performance in terms of throughput achieved and the percentage of dropped packets to achieve this throughput. Chain throughput demonstrates the amount of traffic successfully transferred per unit time where as packet drops determine how efficiently this traffic was transferred.
\begin{figure}[htp]
\begin{center}
        \subfigure[Throughput.\label{fig:chaintp}]{\includegraphics[scale=0.27]{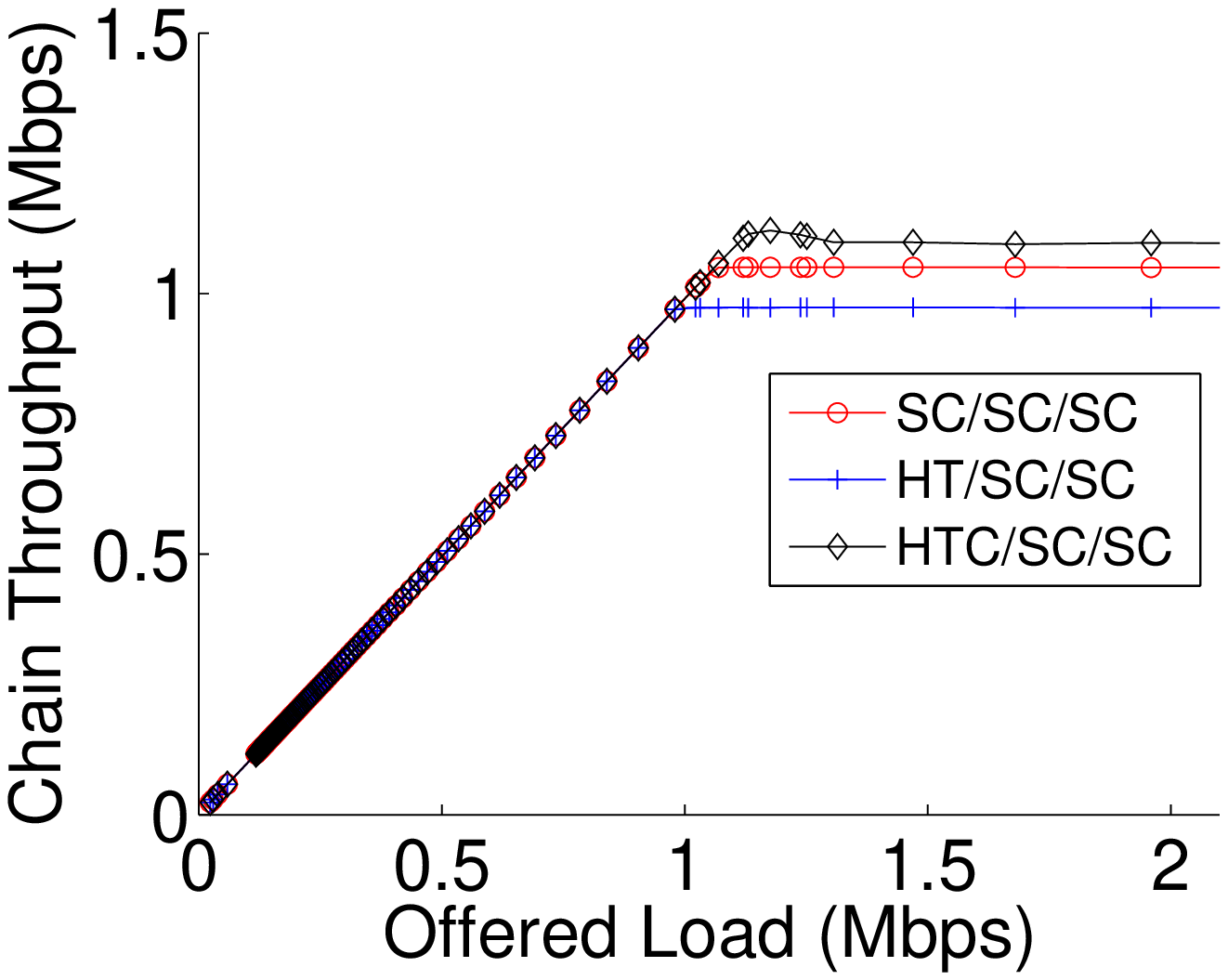}} 
        \subfigure[Percentage of packets dropped.\label{fig:chainbw}]{\includegraphics[scale=0.27]{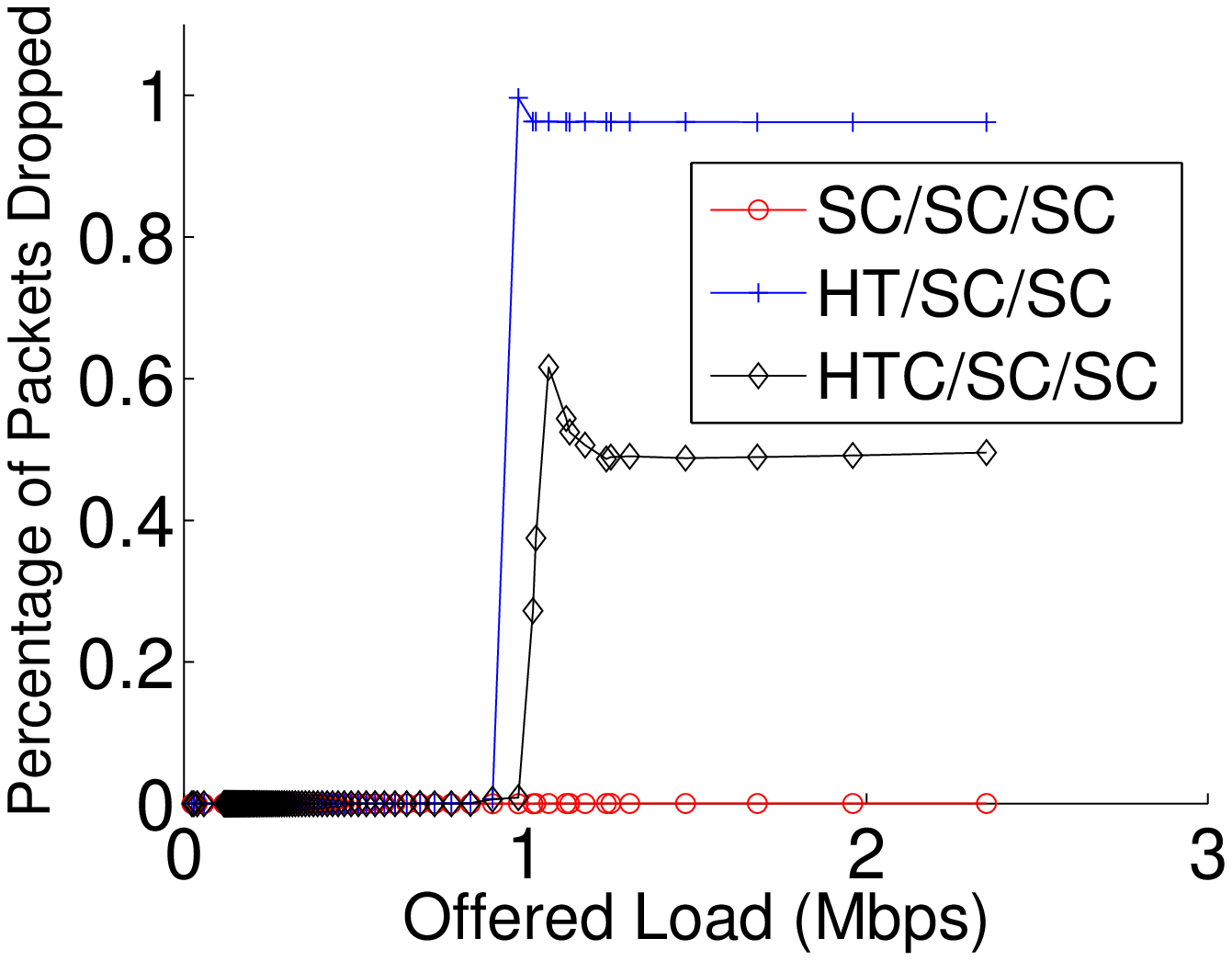}} 
\end{center}
\caption{Simulation based Performance Analysis of 4-hop Chains vs Channel Saturation.}
\label{fig:chaintpall}
\end{figure}

\begin{figure}[htp]
\begin{center}
        \subfigure[Throughput.\label{fig:chaintptb}]{\includegraphics[scale=0.27]{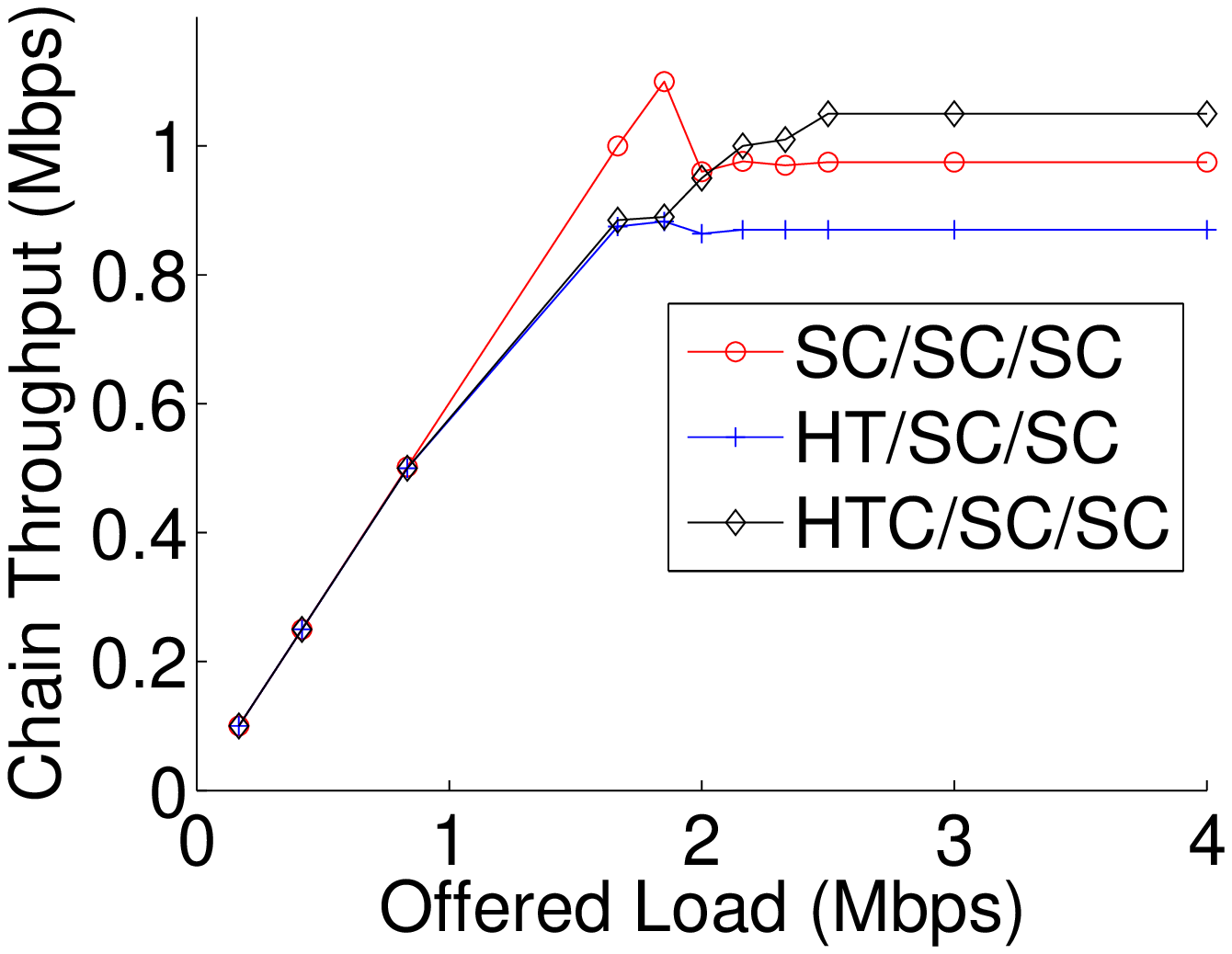}} 
        \subfigure[Percentage of packets dropped. \label{fig:chainbwtb}]{\includegraphics[scale=0.27]{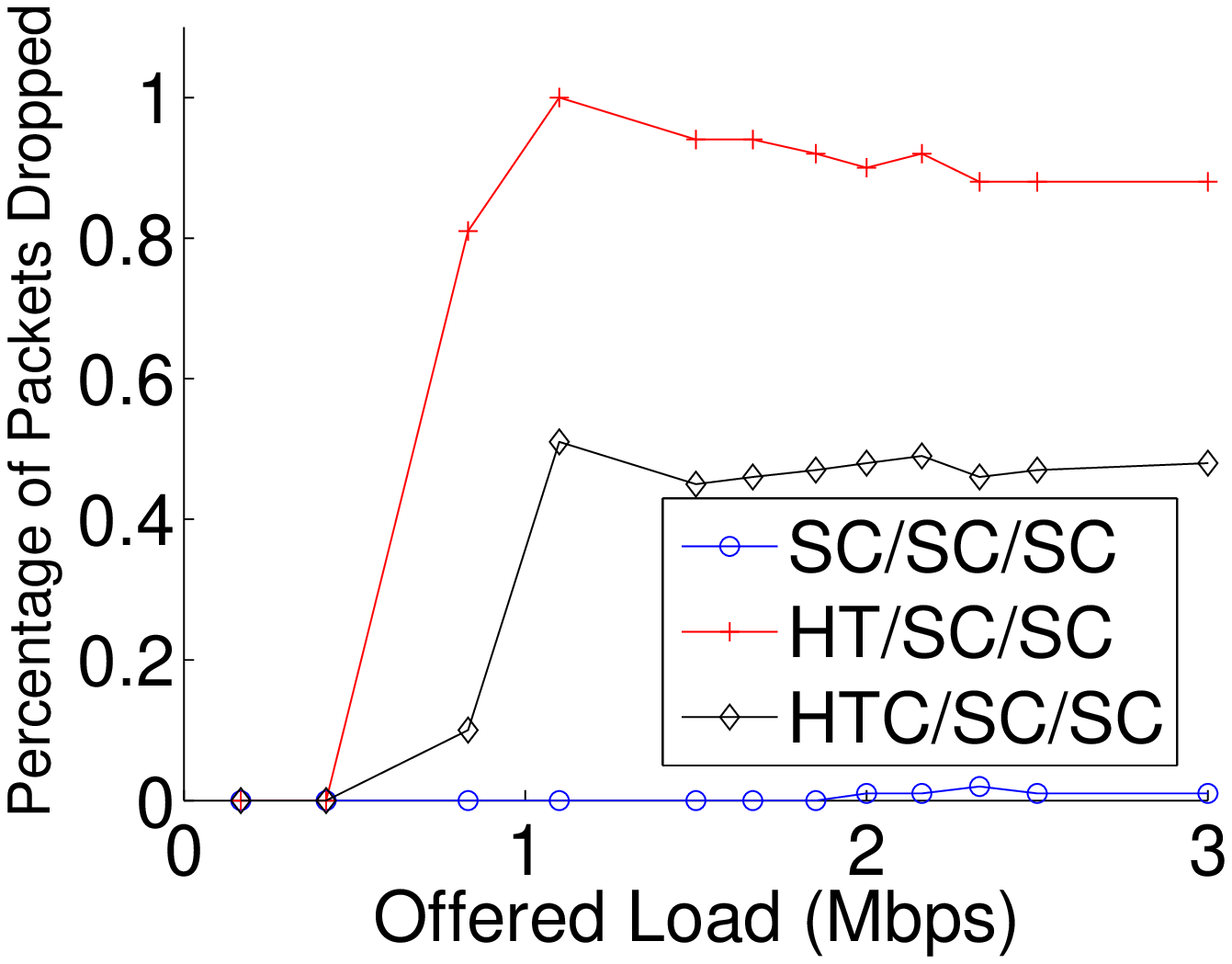}} 
\end{center}
\caption{Performance Analysis of 4-hop Chains on Wireless Testbed.}
\label{fig:chainalltestbed}
\end{figure}

We first carry out simulation based studies using Network Simulator (NS2) ~\cite{ref:ns2} to study the performance of these chains in an environment with repeatable results. We then conduct the same experiments in a wireless testbed in order to study the accuracy of our results in a more realistic environment. 

\subsection {Simulation Based Performance Analysis}
We simulate scenarios with 4-hop routes. We
use a fixed distance of 250m for transmission range and disable
RTS/CTS mechanism. All transmissions are based on 802.11 DCF mode at
data rates of 6Mbps and packet size of 1500 bytes. Our choice for 6Mbps was based on our testbed evaluation. In our testbed since we are using 802.11a to get interference free channels the minimum supported rate is 6Mbps. We change the saturation level of the channel by altering the rates at which the
source pumps Constant Bit Rate (CBR) traffic into the chain.  We
perform this analysis using the standard two-ray ground wireless
propagation model with SINR model for packet reception and capture effects. We fix the Carrier Sense range at 550 meters. 

Figure~\ref{fig:chaintp} shows that for all interactions, chains behave similarly at low saturation levels. Saturation levels determine how often a source transmits a packet; at full saturation, the sender will always have a packet to transmit. At  low levels of saturation, a packet transmitted from the source makes it to the destination before the next packet is transmitted at the source. Hence, there is no interference between the links. The reason for the sudden jump in packet drops for HT cases is that as soon as we cross the saturation threshold, a packet sent on the first hop will collide with the transmissions on the last link. That packet will be successfully retransmitted. When this packet is eventually transmitted on the last hop, it causes the new packet being transmitted on the first hop to be dropped. This way each packet will be dropped once on the first link before it is successfully transmitted. Hence, we see almost $100\%$ drops as soon as we cross the saturation threshold.   

As saturation increases, the level of contention between links also increases causing throughput to eventually level-off, we call this the {\em chain effect}. As we see from Figure ~\ref{fig:chaintp}, the chain effect has a higher impact on throughput than self interference interactions since all three interactions show similar performance. 

The interesting observation, however, is that the three chains consume different amounts of channel bandwidth to obtain this same throughput as shown in Figure~\ref{fig:chainbw}. At lower saturation, the behavior of each chain is the same; as none of the chains experience any packet drops. As saturation levels increase, the interactions between links start to affect chain performance. For chain with Hidden Terminals, more packets are dropped resulting in extra bandwidth usage. The reason why these packet drops do not substantially effect the throughput of the chain is that these packets in Hidden Terminal (HT) chains are transmitted at times when Sender Connected (SC) chains are waiting to transmit because of channel contention. Hence waiting periods in Sender Connected chain are used to transmit packets that are dropped in hidden terminal chains. Percentage of packets dropped in Hidden Terminal with Capture (HTC) cases is better than HT cases, as several packets sent by the source are captured on the first link.


\subsection {Testbed Evaluation}
\begin{figure}[htp]
 \centering
\includegraphics[scale=0.32]{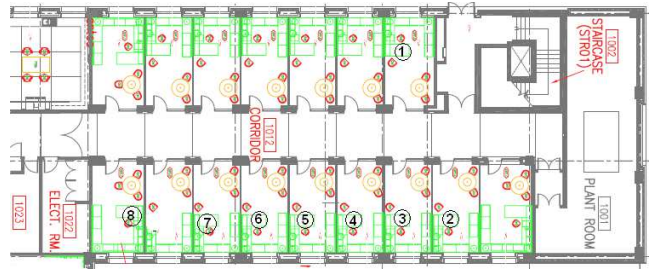}
\caption{Testbed layout in the computer science building.}
\label{fig:testbedmap}
\end{figure}

We validate the results for chain performance obtained from simulation
using a wireless testbed to confirm whether our observations will hold
in a real network. Our testbed consists of 8 nodes that are placed in
offices on the same floor of our building as shown in
Figure~\ref{fig:testbedmap}

Each numbered circle represents a single wireless node. Each node
consists of a soekris board~\cite{ref:soekris_4826} with mini-PCMCIA
wireless card running atheros chipset~\cite{ref:atheros} and madwifi
device driver~\cite{ref:madwifi}. We operate the wireless cards on
802.11a to avoid interference with our resident 802.11 b/g
network. There is also an 802.11a network inside the building but it
uses four channels at the lower end of the spectrum so we set our
wireless cards to operate on channel 157 (5.785GHz).  We use a bit
rate of 6Mbps and disable RTS/CTS handshake.

We start each node and check the kind of interactions that occur
between nodes. Nodes 3 and 7 can sense each others transmission so
they are Sender Connected (SC). To make sure that links 3-4 and 7-8
are Sender Connected (SC), we start transmission from node 3 to node 4
and from node 7 and node 8. We observe that both links equally share
the channel. We create a chain by adding static routes so that packets
being sent from Node 3 to Node 8 are routed through nodes 4, 6 and 7
making a four-hop route 3-4, 4-6, 6-7, and 7-8. This creates an
SC/SC/SC chain since all nodes are in Carrier Sense range.

To create an HT/SC/SC interaction we pick the source to be node 1 and
destination to be node 8. The 4-hop route between these two nodes goes
through nodes 3, 5, and 7. Nodes 1 and 7 are out of range so they can
transmit together. When node 7 transmits, node 3 is unable to capture
packets from node 1 creating a hidden terminal interaction between
links 1-3 and 7-8. To verify this interaction, we start transmitting
packets from node 1 to node 3 on link 1-3. We see the maximum
throughput on this link. Now we start transmitting packets from node 7
to node 8 on link 7-8. The throughput on link 1-3 drops substantially
and we see most of the packets on this link being
dropped~\cite{ref:srazak_twoflow}. Hence chain 1-3-5-7-8 represents an
HT/SC/SC interaction.

To create an HTC/SC/SC interaction we use the same chain that gave us
an HT/SC/SC interaction: 1-3-5-7-8 and start reducing transmission
power on node 7 until node 3 is able to capture packets from node 1.
Figure~\ref{fig:chainalltestbed} show results obtained from the
testbed. We see that the results closely match those obtained from
simulation although the throughput from testbed is slightly lower than
simulation. We attribute this to the physical hardware delays in the
real network.

Current routing protocols do not consider the interaction between
links of a chain while making routing decisions. Metrics that try to
maximize throughput of a route will consider chains with these
different interactions to be similar and will pick routes irrespective
of their efficiency. This will cause suboptimal usage of network
bandwidth; an already limited resource, hence causing lower throughput
in the whole network. Inefficient routes also require more transmits
for each successful transmission, which also wastes the limited energy
resources of wireless nodes.


\section{Generalization to n-hops}
\label{sec:nhops}
In this section we use our results from 4-hop chains to generalize to
n-hop chains. In a chain, each hop can possibly interact with every
other hop within the chain. Therefore in an n-hop chain, the first hop
will interact with $n-1$ hops, second hop will interact with the
subsequent $n-2$ hops, and so on. The total number of interactions
$N_i$, between hops is thus given by the following equation:
\begin{equation}
 N_i = \frac{n(n-1)}{2}
\end{equation}
Previously, we had determined that there are 10 different types of
interactions between two flows~\cite{ref:twoflow_sinr}. Consequently
it is possible to have each of the interactions to be one of the 10
states. This makes the total number of possible interactions between
hops of a chain to be $10^{N_i}$. Clearly this is an intractable
number to analyze. We have observed in our evaluation of 4-hop
interactions that out of the 10 interactions possible, 3 occur most
frequently in chains because of the their geometric nature. Hence we
can reduce the number of considered interactions to $3^{N_i}$.

To further reduce the $N_i$ term, we make the following observations:
\begin{itemize}
 \item The destination of each hop is the source of the next
   subsequent hop. Since nodes within a hop are always within
   communication range of each other, each hop has an SC interaction
   with its neighboring hop.
 \item For commonly used values of Carrier Sense ranges in commercial
   radios (equal to two times the Communication Range or more) two
   hops separated by a single hop are always going to have SC
   interaction as well. The reason for this is that the sources of
   these two hops share a common neighbor i.e. the source of the
   middle hop, and since neighbors can be at most Communication Range
   apart. Therefore, the distance between the source of two hops
   separated by a single hop can be at most two times the
   Communication Range.
 \item All chains start from a source and go towards a
   destination. Hence each hop of the chain goes further away from the
   source and gets closer to the destination. This causes enough
   distance between links such that if there are enough hops between
   two links, the links will not have any interaction between them. By
   analyzing routes with more than 4 hops we observe that in more than
   99\% cases, there is no interaction between links that are 3 hops
   apart using the NADV forwarding rule.
\end{itemize}

From the above observations we make the approximation that a hop $a$
will have SC interactions with hops $a+1$ and $a+2$, one of the three
interactions with hop $a+3$ and no interaction with any subsequent
hops. This simplification reduces the total number of interactions to
$n-3$, since the last three hops dont have an $a+3$ neighbor to
interact with.  Hence an n-hop chain can have a total of $3^{n-3}$
interactions.

To determine the behavior of n-hop chains, we make the following observations based on our results:
\begin{itemize}
 \item The throughput of a chain is independent of the type and number of self-interfering interactions.
 \item In contrast, the performance of a chain in terms of number of
   dropped packets, depends upon the types of interfering interactions
   between links of the chain as well as the location of these
   interactions. The kind of interaction towards the beginning of a
   chain will have a higher impact on the performance than the later
   interactions.
 \item Chains with SC interactions perform better than those with HTC
   or HT interactions.  Chains with HTC interactions perform better
   than those with HT interactions.
\end{itemize}

\begin{figure}[htp]
\begin{center}
        \subfigure[Throughput.\label{fig:tp5hops}]{\includegraphics[scale=0.27]{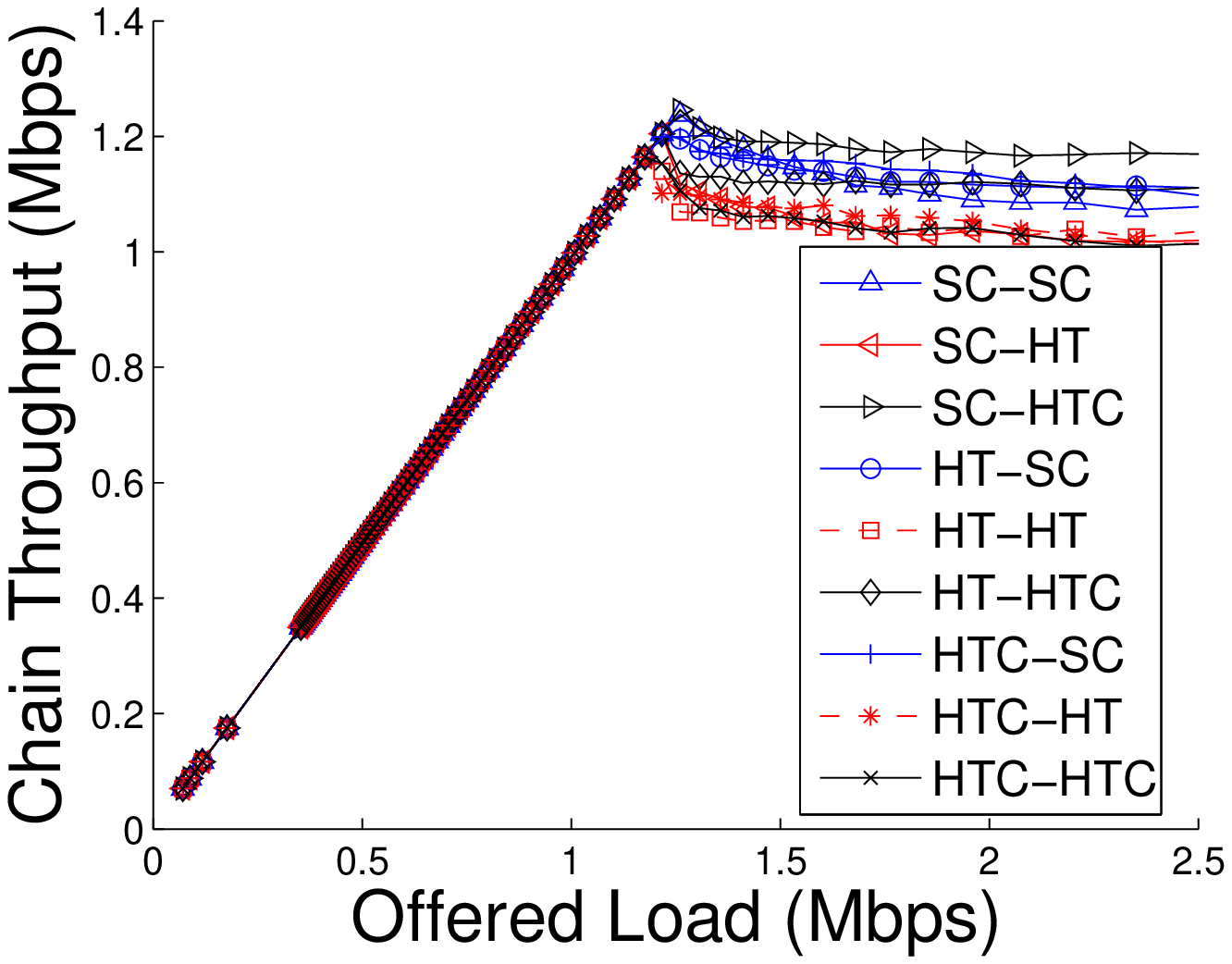}} 
        \subfigure[Percentage of packets dropped.\label{fig:bw5hops}]{\includegraphics[scale=0.27]{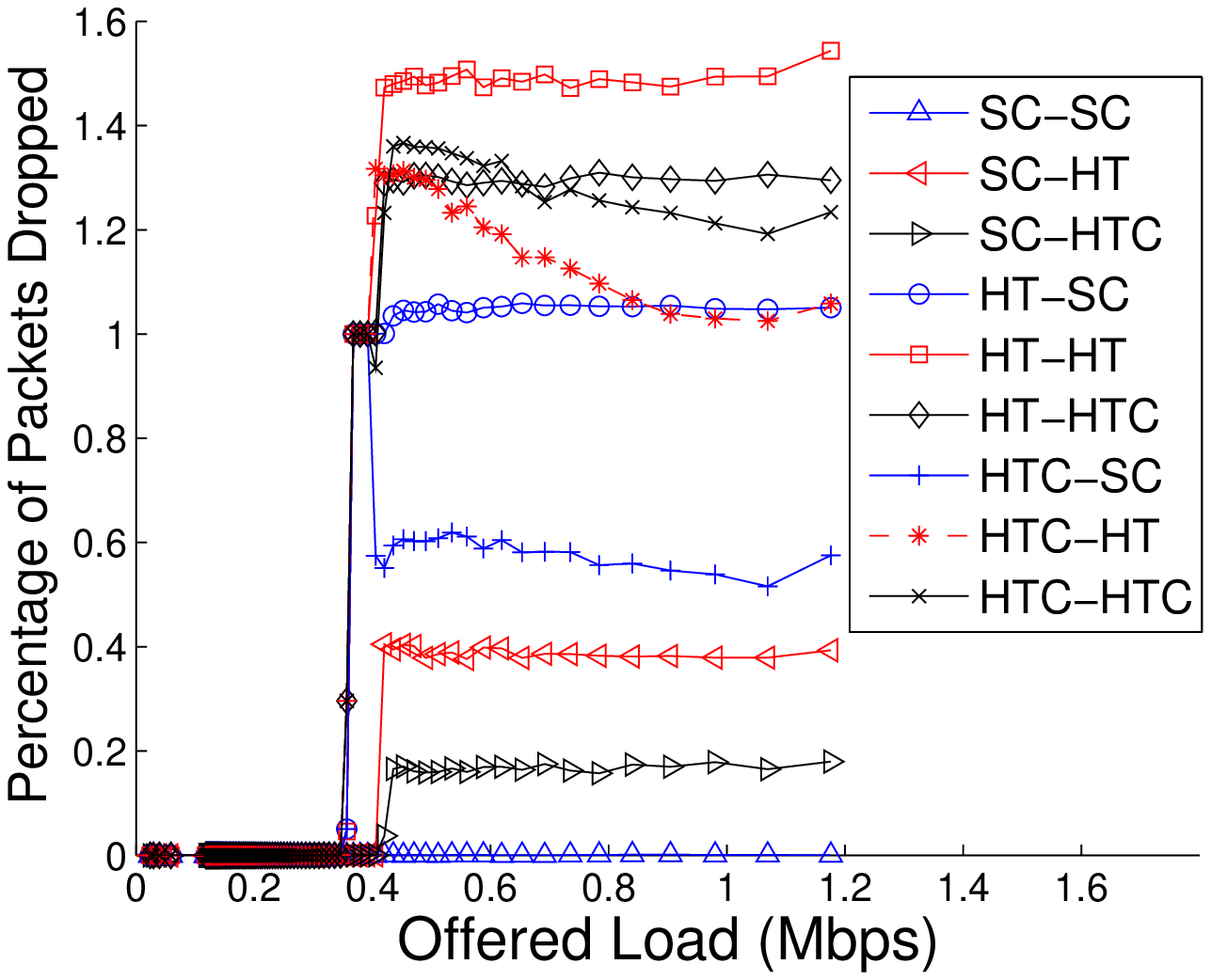}} 
\end{center}
\caption{Performance Analysis of 5-hop Chains with Different Interaction Combinations: HT interactions cause worst performance in terms of packet drops followed by HTC and then SC. Throughput is mostly independent of interference interaction.}
\label{fig:chain5hops}
\end{figure}

Figure~\ref{fig:chain5hops} illustrates the performance of a 5-hop
chain, with different link1-link4 and link2-link5 interactions. In the
figure, the terminology SC-HT means that there is an SC interaction
between the first and fourth hops and an HT interaction between the
second and fifth hops. The plot shows that the difference between the
best and worst throughput is less than $15\%$. We also observe that SC
interactions perform better, especially when they occur at the
beginning of the chain. HT interactions at the start of the chain have
the worst performance.

\begin{figure}[htp]
\begin{center}
        \subfigure[Throughput.\label{fig:tp8hops}]{\includegraphics[scale=0.27]{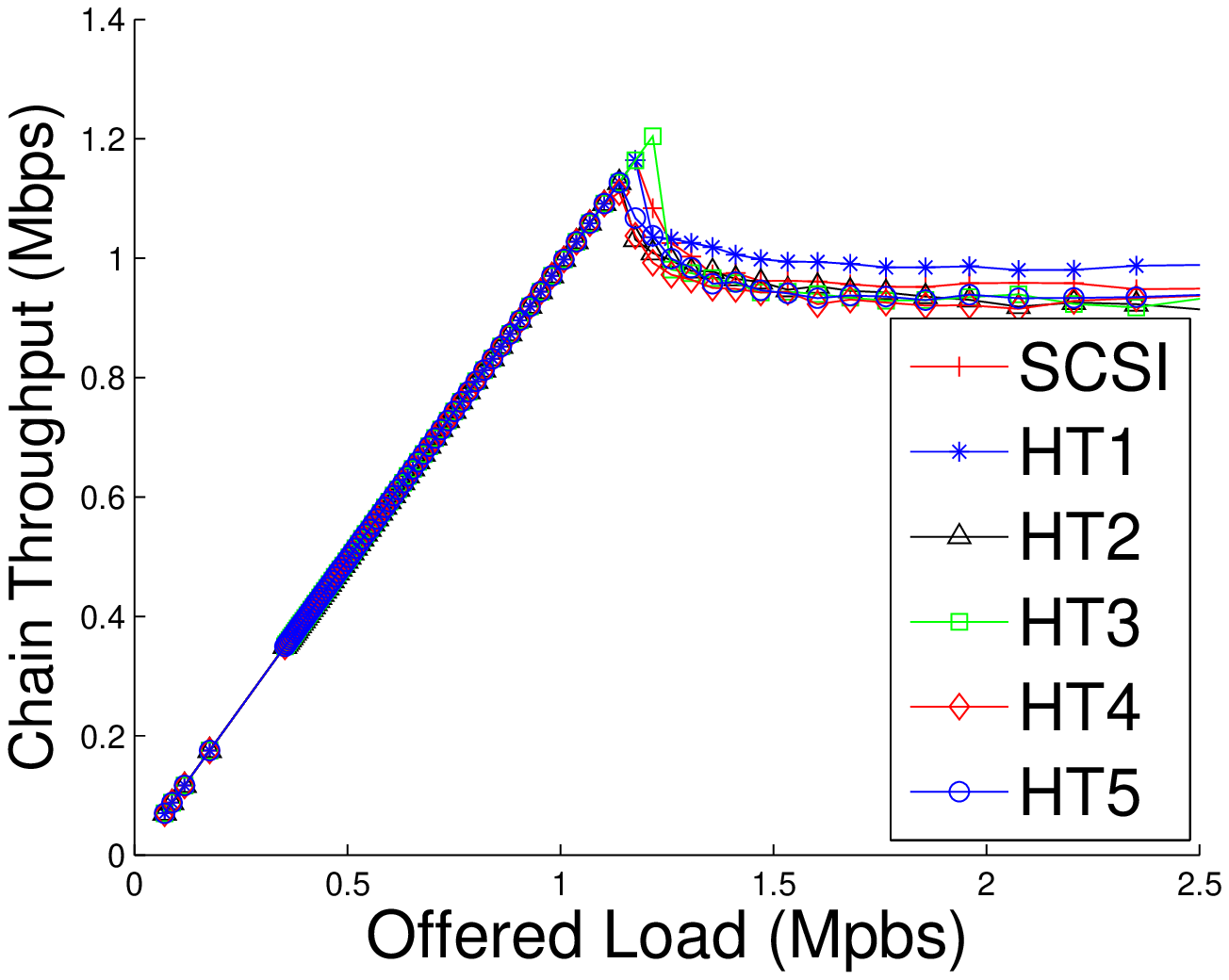}} 
        \subfigure[Percentage of packets dropped.\label{fig:bw8hops}]{\includegraphics[scale=0.27]{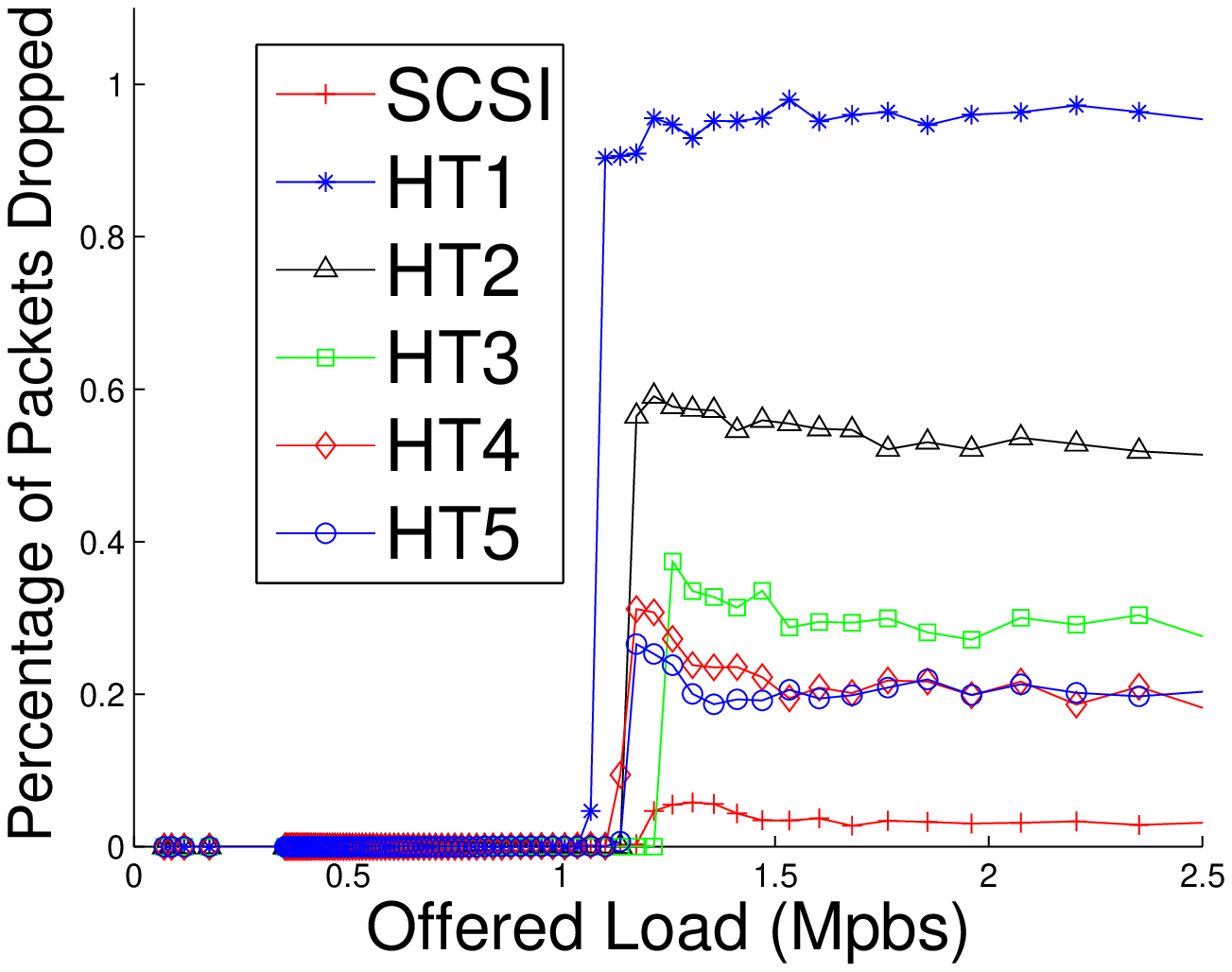}} 
\end{center}
\caption{Impact of HT location in 8 hop chain: HT interactions at the beginning of the chain cause more packet drops than later interactions, while the throughput overall stays unaffected.}
\label{fig:chain8hops}
\end{figure}
Figure~\ref{fig:chain8hops} illustrates the performance of 8-hop
chains that have HT interactions at varying distances from the
beginning of the chain. We observe that HT interactions at the
beginning of the chain are much more pronounced than in later hops,
with the effect being minimized after the chain throughput stabilizes
after the first few hops.  These observations allow us to analyze the
first few interactions in a chain and compare their
performance. Quantifying the effect of these interactions and applying
them as metrics for routing is part of our future work.



\section{Interactions across chains}
\label{sec:crosschain}

This section considers the problem of interactions that occur across
different chains.  Clearly, the number of possible interactions that
general chains can have with each other is overwhelming, preventing a
systematic analysis such as the one we attempted with a single chain.
Instead, the study presented in this section examines the following
questions: (1) Does the type of chain (from a self-interference
perspective) affect its susceptibility to cross chain interference?
(2) Do different types of chains interact differently with each other?
(3) What is the effect of cross chain hidden terminals and hidden
terminals with capture on different types of chains?  

For simplicity, we denote the SC/SC/SC, HTC/SC/SC and HT/SC/SC chain
categories as SC, HTC and HT respectively. We simulate cross-chain
interactions by randomly choosing two chains of a particular category
that were selected by the NADV forwarding rule. This approach leads to
6 possible combination of cross-chain interactions: 2 SC chains, 2 HTC
chains, 2 HT chains, SC \& HTC chains, SC \& HT chains and HTC \& HT
chains. We analyzed more than 200 different scenarios under each
category.
 We analyze the number of cross-chain interactions that occur when two
chains interact. We then study the effect of hidden terminal
interactions on different combinations of chains.
\begin{figure*}
\begin{center}
	\mbox{
	\subfigure[Occurrence probability of cross-chain hidden terminals in chains: A significant number of hidden terminals emerge during cross-chain interactions. HT and HTC chains are more vulnerable to hidden terminals than SC chains. \label{fig:randomChains.rand.htInChains}]{\includegraphics[scale=0.5]{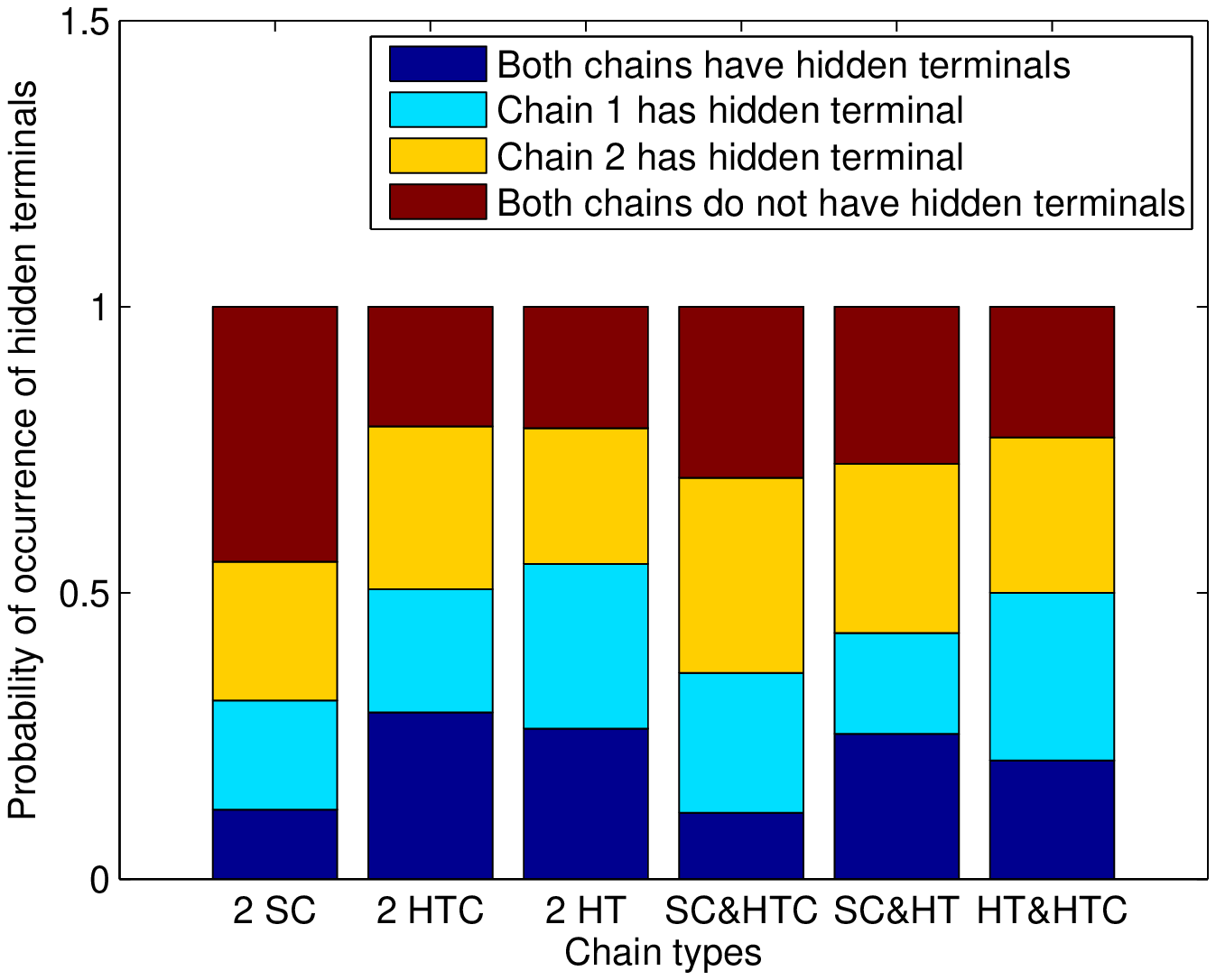}}\quad
	\subfigure[Effect of hidden terminals on throughput in some representative scenarios: Severe unfairness results when hidden terminal is present in a HT or HTC chain, but absent in a competing SC chain. \label{fig:randomChains.rand.repTh}]{\includegraphics[scale=0.5]{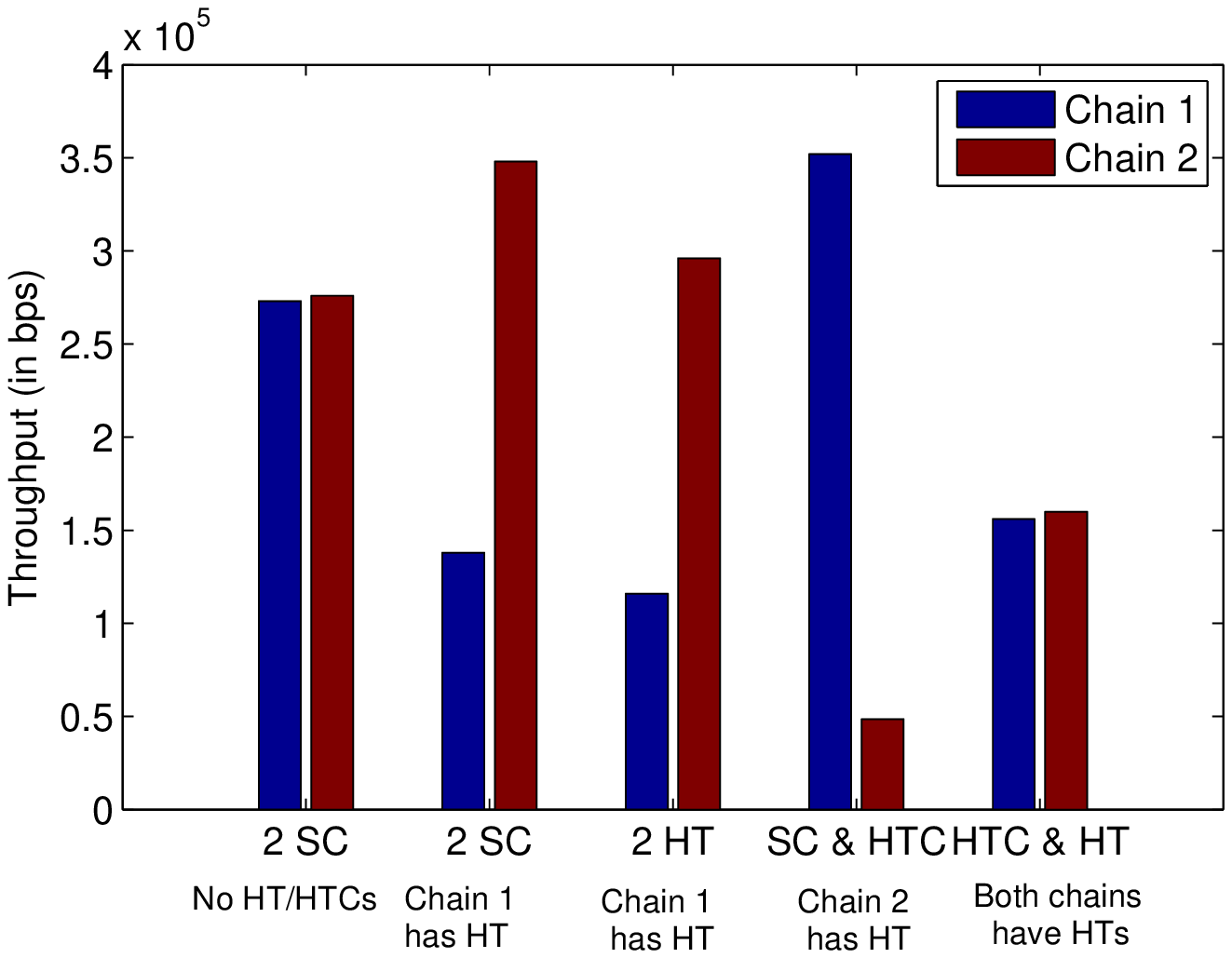}}
	}
\end{center}
\caption{Hidden terminals in cross-chain interactions.}
\label{fig:randomChains.interactions.res}
\end{figure*}

\paragraph*{Cross-chain hidden terminals}

We empirically study the occurrence probability of a hidden terminal
on two chains. Figure~\ref{fig:randomChains.rand.htInChains} shows
occurrence probabilities of a hidden terminal (HT or HTC interaction)
in two chain scenarios. We observe that the occurrence of Symmetric HT
(or Symmetric HTC), where a pair of links have HT (or HTC) to each
other, are very rare. We refer to the first and second chains in each
interaction category as \textit{chain 1} and \textit{chain 2},
respectively. Figure~\ref{fig:randomChains.rand.htInChains} shows that
weaker chains like HT or HTC chains are more vulnerable to cross-chain
hidden terminal interactions than the SC chain. It can be observed
that the probability of occurrence of hidden terminal in at least one
of the chains is very high: a value of $0.55$ in the 2 SC chains and
$0.8$ in weaker chains.  The reason that HT and HTC chains suffer more
cross chain hidden terminals is that the link with a HT in
self-interference has low SINR at the receiver, meaning that it is
susceptible to interference from another chain as well.  The average
throughput values of representative scenarios are shown in
Figure~\ref{fig:randomChains.rand.repTh}. 

We illustrate the general effects of hidden terminals through
representative scenarios due to space limitations.  In general, severe
unfairness from hidden terminals result when a strong chain like SC
interacts with a weaker ones like HTC and HT chains. Weaker chains
with cross-chain hidden terminal interactions are prone to severe
throughput degradation. This is highlighted when HT \& HTC chains
interact and both have cross-chain hidden terminals: the presence of
hidden terminals reduces the throughput of both chains to
approximately half the original value. The amount of throughput
degradation is dependent upon the placement of the two links that are
involved in the hidden terminal and the type of self-interference
observed. The average throughput of two interacting chains are equal
when two similar chains have no hidden terminals. Even under such
stable scenarios, we observed a large variation of throughput due to
contention unfairness. We move on to explain this effect.
%

\paragraph*{Effect of contention unfairness}
We now examine a collective interaction that significantly affects the
performance of the links. Even in the absence of hidden terminals,
some links may suffer starvation due to very low channel access
probabilities due to contention from other links. 
The contention unfairness problem problem can be explained by a simplified topology that
 highlights its effect.
 \begin{figure}
 \centering
\includegraphics[scale=0.3]{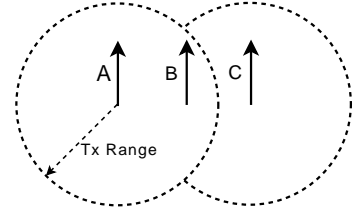}
\caption{Flow in the Middle: Representative example for contention unfairness.}
 \label{fig:fim}
 \end{figure}
In the Flow-in-the-middle topology shown in Figure~\ref{fig:fim}, Link
A and C do not sense each other and can transmit in parallel. However,
link B can only transmit when neither A \textit{nor} C are
transmitting. Due to the fact that A and C are out of range, and
unsynchronized, B experiences large busy times on the channel, as the
channel is mostly busy with transmissions from either A or C (or
both).  This leads to severe starvation. The other links capture a
large share of transmission time. We call such interactions
``contention unfairness'' since individual links experience unfairly
different contention levels.

 The pipeline effect of traffic on a single chain reduces the impact
 of contention unfairness.  However, when independent chains compete,
 contention unfairness may arise, leading to queue drops and
 unfairness.



\paragraph*{Combined impact on chain performance}
\begin{figure}
 \centering
\includegraphics[scale=0.5]{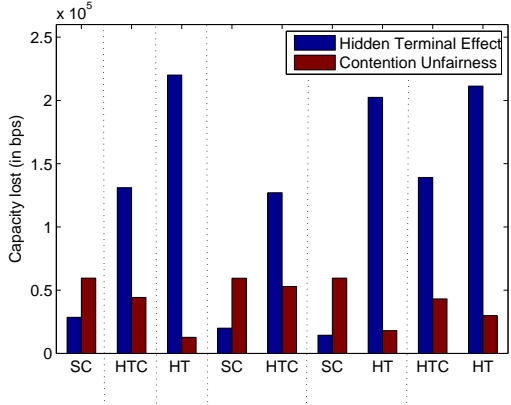}
\caption{Capacity wasted due to hidden terminals and contention unfairness: The effect of contention unfairness is significant in cross-chain interactions.}
\label{fig:randomChains.rand.htContention}
\end{figure}
Figure~\ref{fig:randomChains.rand.htContention} compares the effect of
hidden terminals and contention unfairness in terms of the lost
capacity. Contention unfairness significantly reduces capacity for all
chains.  The large variance indicates that the effect of hidden
terminals and contention unfairness cannot be accurately predicted in
the aggregate; rather, a case-by-case analysis is required.

\paragraph*{Chain type and vulnerability to cross chain interactions}
The next experiment studies the vulnerability of the different chain
types to destructive cross-chain interactions.  Considering the types
of interactions in order of severity (HT, HTC followed by SC), we
label a link in a chain according to the most severe interaction it
suffers.  For example, if a link has an HT interaction from one link,
HTC from another link and SC with some others, the link is labeled HT.
We then empirically calculate the conditional probability that the
link has an interaction $X$ from the other chain given that it had
interaction $Y$. This metric quantifies the vulnerability to
cross-chain interactions for the link with $X$ interaction under
self-interference.
\begin{figure}[ht]
\begin{center}
\includegraphics[scale=0.5]{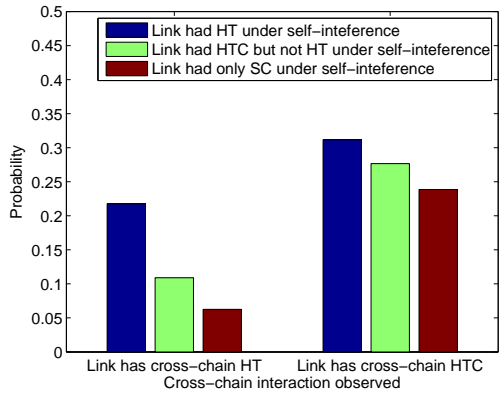}
\end{center}
\caption{Conditional interaction probability: Weak links (HT and HTC) under self-interference have high probability of being weak in cross-chain interactions. Links that have only SC interactions under self-interference are much more stable.}
\label{fig:rand.conditionalInteractionProb}
\end{figure}

Figure~\ref{fig:rand.conditionalInteractionProb} shows this
conditional interaction probability. It can be seen that weak links
(ones having HT or HTC) have much greater probability to have
detrimental interactions than the links that have only SC
interactions.
\section{Conclusions and Future Work}
\label{sec:conclusion}

Chains are a fundamental communication structure in multi-hop wireless
networks; understanding their performance is important for building
efficient protocols.  The behavior of chains is complicated because of
a number of complex processes that arise as different hops of a
chain interfere with each other.  We first identify these different
effects and argue that MAC level interactions are the most important.
We analyze the frequency of occurrence of the different types of
chains, classified by the types of MAC interactions they exhibit.
There is a large number chain types possible in four hop chains when
we consider the different interaction combinations that can
arise. However, the analysis shows that only a small number of these
interactions occurs in practice.  

We also study the performance of chain types that occur.  We find that,
even though the different chains show similar throughput when
considered in isolation, different chains experience significantly
different number of packet drops.  These packet drops, require
retransmissions, increasing the resources required to forward the
packet and harming the overall performance of the network.  We
validate these results using testbed experiments.
We also generalize the results we obtained four-hop chains to
n-hop chains.  

The paper then analyzes the link level interactions that occur across
two interfering chains.  The number of potential interactions grows
exponentially, preventing a systemic analysis.  However, we make a
number of interesting observations: (1) MAC level interactions also
play a primary role in how multiple chains interaction.  Other factors
such as contention unfairness play a smaller role.  For example, the
presence hidden terminals (with or without capture) across chains
significantly hinders performance and fairness; and (2) A chain that
is well behaved with respect to self-interference is more immune to
interference from another chain.

Our future plans include deeper analysis of cross chain interactions.
We also plan to develop routing protocols that leverage the
observations made about chain behavior.  Another intriguing
possibility is to explore changing MAC parameters such as transmission
power and carrier sense range to change chains that have destructive
interference into better chains that do not.

%
\bibliographystyle{IEEEtran}
\bibliography{references}  
%
\end{document}